\documentclass[a4, 10 pt, conference]{ieeeconf}  % Comment this line out if you need a4paper

\IEEEoverridecommandlockouts                              % This command is only needed if 
\usepackage{verbatim}
\usepackage{url}
\usepackage{graphicx}
\usepackage{amsmath, amssymb}

\title{\LARGE \bf
Team OS's System for Dialogue Robot Competition 2022
}

\author{Yuki Kubo, Ryo Yanagimoto, Hayato Futase, Mikio Nakano, Zhaojie Luo, Kazunori Komatani% <-this % stops a space
\thanks{The~Institute~of~Scientific~and~Industrial~Research~(SANKEN), 
	Osaka~University, Osaka, Japan \{y-kubo, r-yanagimoto, h-futase\}@ei.sanken.osaka-u.ac.jp, mikio.nakano@c4a.jp, \{luo, komatani\}@sanken.osaka-u.ac.jp}%
}

\begin{document}

\maketitle
\thispagestyle{empty}
\pagestyle{empty}

%%%%%%%%%%%%%%%%%%%%%%%%%%%%%%%%%%%%%%%%%%%%%%%%%%%%%%%%%%%%%%%%%%%%%%%%%%%%%%%%
\begin{abstract}
% Our dialogue robot system, OSbot, developed for Dialogue Robot Competition 2022 is introduced.
This paper describes our dialogue robot system, OSbot, developed for Dialogue Robot Competition 2022.
The dialogue flow is based on state transitions described manually and the transition conditions use the results of keyword extraction and sentiment analysis.
The transitions can be easily viewed and edited by managing them on a spreadsheet.
The keyword extraction is based on named entity extraction and our predefined keyword set.
The sentiment analysis is text-based and uses SVM, which was trained with the multimodal dialogue corpus Hazumi. 
We quickly checked and edited a dialogue flow by using a logging function.
% During the development, we quickly checked dialogue logs generated by our testing function and edited a dialogue flow.
% During the development, we carefully checked dialogue logs generated by our testing function.
In the competition's preliminary round, our system ended up in third place.

\end{abstract}

\section{Introduction}
We developed a dialogue robot system named OSbot for Dialogue Robot Competition 2022 \cite{IEEEref:robotcompe_2022}.
OSbot integrates the multimodal input/output modules distributed by the competition organizers \cite{IEEEref:robotcompe_robot} and our developed dialogue processing module. 
The dialogue processing module is based on DialBB\footnote{DialBB was developed by C4A Research Institute, Inc. and released for non-commercial usage at https://github.com/c4a-ri/dialbb}, a framework for dialogue system development.  
OSbot adopts a dialogue strategy similar to that of Team kasuga's system, which won second place in the past Dialogue Robot Competition \cite{IEEEref:robotcompe1}, and reuses some of its modules.
% OSbot adopts a dialogue strategy similar to that of Team kasuga's system, which won second place in  Dialogue Robot Competition 2021 \cite{IEEEref:robotcompe_1}, and reuses some of its modules.

Below are the characteristic features of OSbot:
\begin{itemize}
    \item OSbot uses state transition network-based dialogue management to explicitly control the dialogue flow and avoid unexpected system utterances that do not match the context at all.
    \item OSbot uses the results of keyword extraction and sentiment analysis for the conditions of state transition.
    The results of keyword extraction are also used in system utterances.
\end{itemize}

In the following, we explain the architecture of OSbot and the ideas behind it and show the results obtained in the preliminary round.

\section{OSbot System}
 In this section, we explain the module components and module integration of our proposed OSbot system.   

\begin{figure*}[ht]
  \centering
  \includegraphics[width=15.5cm]{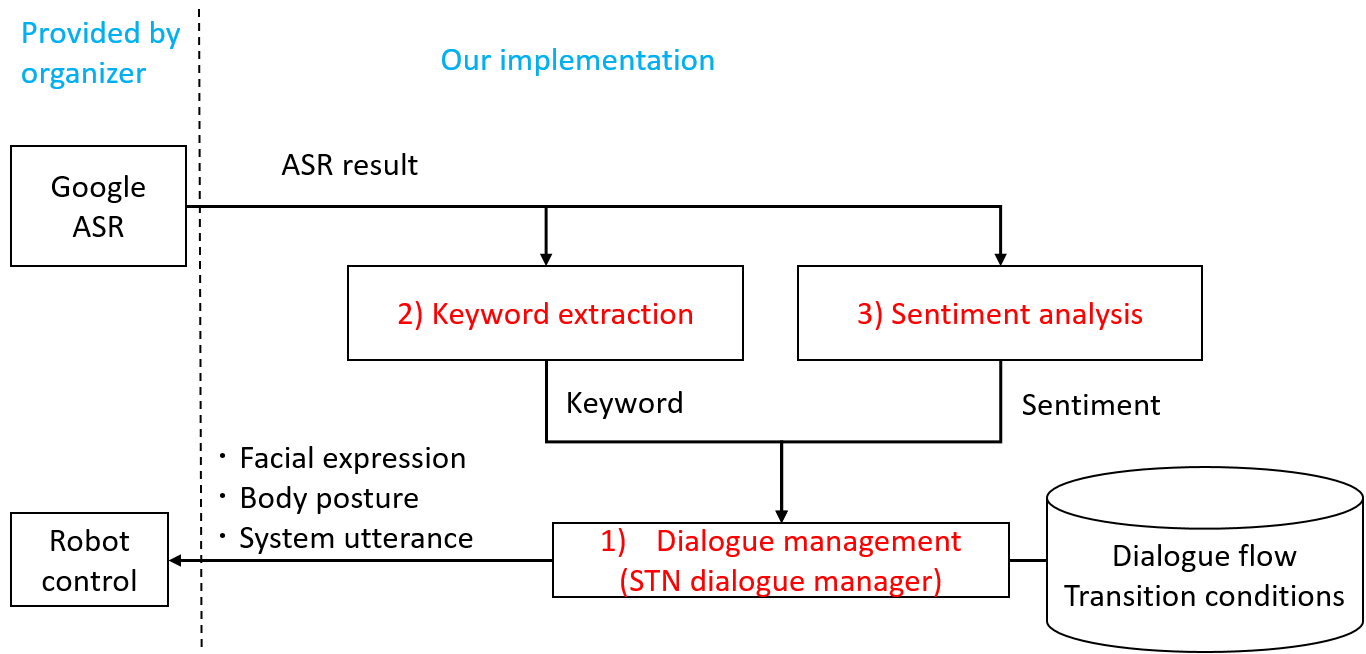}
  \caption{
  Module architecture of OSbot. Right-hand side of dotted line is OSbot's dialogue processor.
  }
  \label{fig:sys_whole}
\end{figure*}

\subsection{Module components}
As shown in Fig.~\ref{fig:sys_whole}, our OSbot system consists of three important modules: 1) dialogue management module, 2) sentiment analysis module, and 3) keyword extraction module.

\subsubsection{Dialogue management}
\begin{figure*}[t]
  \centering
  \includegraphics[width=16cm]{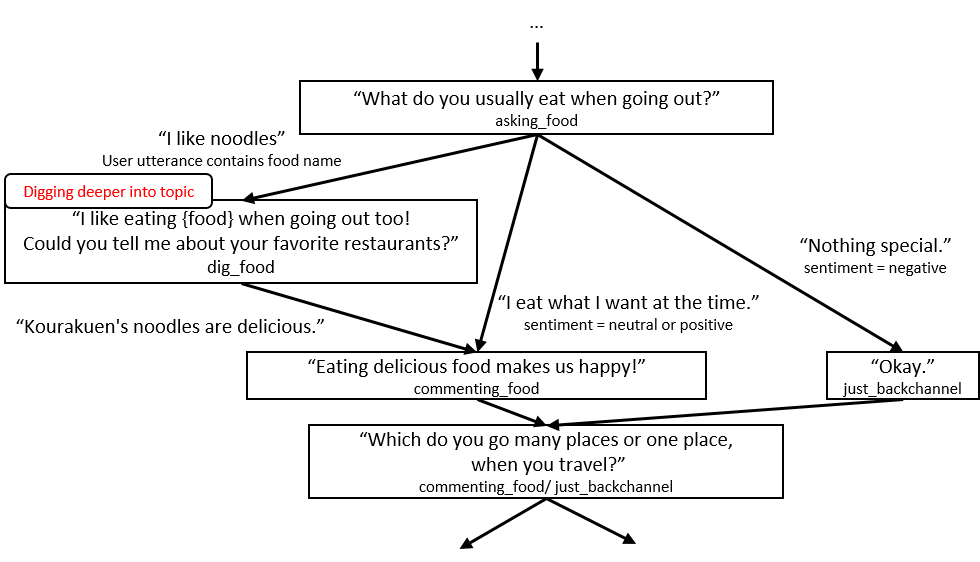}
  \caption{
  Example dialogue flow. Boxes show system utterances. 
  Text near arrows shows example user utterances and conditions for state transitions. 
  }
  \label{fig:dialogue_flow}
\end{figure*}
The dialogue manager uses a hand-crafted state transition network. 
It determines a system utterance on the basis of the state at that point in time. 
It transitions its state to the next state using the results of keyword extraction and sentiment analysis.
It also stores the keyword extraction results specified by the state transition network. 
The stored keywords are used in generating system utterance and in deciding the reasons for recommending a travel destination.
% The stored keywords are used in system utterance generation and in deciding the reasons for recommending a travel destination.

Fig.~\ref{fig:dialogue_flow} shows an example dialogue flow.
The dialogue manager uses the results of keyword extraction and sentiment analysis to determine the state to which to transition.
System utterances are generated depending on the state.
% System utterances are generated depending on the new state.
In Fig.~\ref{fig:dialogue_flow}, OSbot asks about a user's eating habits.
OSbot has three patterns for the user's reply.
% OSbot has three varieties of utterance patterns by a user's utterance.
If the user's reply includes a food name, OSbot digs deeper into the topic (highlighted in red).
The part of \{food\} is filled with a keyword extracted from the user's utterance.
If the sentiment analysis outputs neutral or positive, OSbot shows only a comment like ``Eating delicious food makes us happy!''
Otherwise, OSbot shows only backchannels like ``Okay.''
% The part of \{food\} is filled with a keyword extracted from the user's utterance.
% If the sentiment analysis outputs neutral or positive, OSbot shows only a comment like ``Okay.''
% Otherwise, OSbot shows only a comment on the topic.

\subsubsection{Keyword extraction}
Our keyword extraction module uses two methods: named entity labeling using GiNZA\footnote{https://megagonlabs.github.io/ginza/}
% \footnote{\url{https://megagonlabs.github.io/ginza/}}
(natural language processing library) and keyword matching using a predefined keyword set.

First, using GiNZA, the keyword extractor searches for words labeled ``Food,'' ``Dish,'' ``First Name,'' ``Last Name,'' ``Noun,'' and ``Place Name'' in a user utterance.

Second, the keyword extractor searches for keywords in a predefined keyword set. 
It also tries to find a word that matches or is similar to a keyword in the predefined list.
It measures the similarity by cosine similarity of word vectors.
We used a CBOW model of Word2Vec \cite{IEEEref:w2v}, which was trained using the Wikipedia corpus and made public by the Shiroyagi Corporation\footnote{https://github.com/shiroyagicorp/japanese-word2vec-model-builder}.
% \footnote{\url{https://github.com/shiroyagicorp/japanese-word2vec-model-builder}}.

\subsubsection{Sentiment analysis}
We constructed a sentiment analyzer using the Hazumi corpus \cite{komatani_hazumi}, which contains multimodal dialogues between a system and a user.
Our sentiment analyzer is based on a Support Vector Machine (SVM). 
Its output is one of the three sentiment classes (positive, neutral, and negative). 
Its input is a 768-dimension vector that is the output of a pre-trained model (Sentence-BERT \cite{IEEEref:bert}) whose input is a user utterance (in text). 
% Its input is a 768-dimension vector that is the output of a pre-trained model (BERT\cite{IEEEref:bert}\footnote{\url{https://huggingface.co/sentence-transformers/stsb-xlm-r-multilingual}}) whose input is a user utterance (in text). 
We used Sentence-BERT only as the feature extractor because we are planning to combine text embedding with features extracted from other modalities such as audio and vision.

For training the SVM, we used the third-person sentiment annotations (annotated by multiple people using 7 levels of values: 1 is very negative, 4 is neutral, and 7 is very positive) in Hazumi. 
The annotations indicate whether a user is enjoying the dialogue or not.
We converted the set of annotations for each utterance into one of the three labels by calculating the average of multiple annotations and using the thresholds 3.5 and 4.5; 
the number of training samples was 7052: 3533 positive, 2741 neutral, and 778 negative.

\subsection{Module integration}
% \subsubsection{DialBB apprication}
To integrate the modules described above, we used DialBB, a framework for building dialogue systems.

A DialBB application takes user utterance texts as input, and then it outputs system utterance texts.
The main module of a DialBB application sequentially exchanges data with a group of modules called ``Building Blocks'' (hereafter referred to as ``blocks''). 
In the configuration file for each application, a DialBB application can specify which blocks to use.

Developers of DialBB applications can use built-in blocks in DialBB and can also implement their blocks. 
They can use the following three built-in blocks to build a dialogue system.

\begin{description}
    \item[Utterance canonicalizer:]\mbox{}\\
    Canonicalizes an input user utterance string (e.g., replacing uppercase letters with lowercase letters.)
    \item[SNIPS language understander:]\mbox{}\\
    Extracts the dialogue act type and slots from a user utterance using SNIPS NLU \cite{IEEEref:coucke2018snips}.
    \item[STN dialogue manager:]\mbox{}\\
    Performs dialogue management using a state-transition network. 
    It can use developer-defined functions for transition conditions and actions to perform after transitions. Those functions can use the language understanding results.
\end{description}

OSbot uses only the Utterance canonicalizer and the STN dialogue manager among the built-in blocks. 
We implemented a new block for keyword extraction and sentiment analysis. 

% \subsubsection{Dialogueflow in spreadsheet}
\begin{figure*}[t]
% {\tiny
% {\tt
\centering
{\scriptsize
\begin{tabular}{|l|l|l|l|l|l|}
\hline
state             & system utterance                                                                                                   & user utterance example         & conditions                                                                 & actions                                                     & next state        \\ \hline\hline
…                 & …                                                                                                                  & …                              & …                                                                          & …                                                           & …                 \\ \hline
asking\_food      & What do you usually eat when going out?                                                                            & I like noodles.                 & \begin{tabular}[c]{@{}l@{}}user utterance\\ contains \{food\}\end{tabular} & \begin{tabular}[c]{@{}l@{}}memorize\\ \{food\}\end{tabular} & dig\_food         \\ \hline
asking\_food      &                                                                                                                    & I eat what I want at the time. & \begin{tabular}[c]{@{}l@{}}sentiment =\\ neutral or positive\end{tabular}  &                                                             & commenting\_food  \\ \hline
asking\_food      &                                                                                                                    & Nothing special.                &                                                                            &                                                             & just\_backchannel \\ \hline
dig\_food         & \begin{tabular}[c]{@{}l@{}}I like eating \{food\} when going out too! \\ Could you tell me about your favorite restaurants?\end{tabular} & ANYTHING                       &                                                                            & …                                                           & commenting\_food  \\ \hline
commenting\_food  & \begin{tabular}[c]{@{}l@{}}Eating delicious food makes us happy!\\ Which do you…, when you travel?\end{tabular}    & …                              & …                                                                          & …                                                           & …                 \\ \hline
…                 & …                                                                                                                  & …                              & …                                                                          & …                                                           & …                 \\ \hline
just\_backchannel & \begin{tabular}[c]{@{}l@{}}Okay.\\ Which do you…, when you travel?\end{tabular}                                    & …                              & …                                                                          & …                                                           & …                 \\ \hline
…                 & …                                                                                                                  & …                              & …                                                                          & …                                                           & …                 \\ \hline
\end{tabular}
}
% }
\caption{Example of description in spreadsheet. In a spreadsheet, a description surrounded by parentheses is blank.}
\label{fig:example_dialbb}
\end{figure*}
The dialogue manager refers to a spreadsheet in which state transitions are described and determines a system utterance.
% We implemented a dialogue flow by describing state transitions in a spreadsheet. 
% The dialogue manager refers to the spreadsheet and determines a system utterance.
Fig.~\ref{fig:example_dialbb} shows an example description of the spreadsheet.
% The spreadsheet consists of six columns: a system's state (State), a system's utterance in the state (System utterance), an example of what a user utters (User utterance example), conditions that determine the next state (Conditions), processes called at the state transition (Actions), and a system's next state (Next state).
The spreadsheet consists of six columns: the system's state (state), the system's utterance in the state (system utterance), an example of what a user utters (user utterance example), the conditions that determine the next state (conditions), the process called at the state transition (actions), and the system's next state (next state).
We described these manually.
% A sentence that corresponds to the system's state is uttered.
% Conditions that correspond to the system's state are referred to from an upper row, and the condition that matches the criteria is referred to.
% The Actions and Next state whose rows are the same as the condition are called, and the state transition occurs.
Based on this description, the system works as follows: 
first, the system utters the nonblank sentence in the system utterance column of the current state.
% First, the utterance in the system utterance column of the current state is generated. 
Then, conditions from the top of the spreadsheet if the conditions for the current state are satisfied or described blank, the actions in that row are performed and the system moves to the next state.
Fig.~\ref{fig:dialogue_flow} shows the dialogue flow that follows the description in Fig.~\ref{fig:example_dialbb}.

% In Fig.~\ref{fig:example_dialbb}, after the dialogue system utters ``What do you usually eat when going out?'' in the state ``asking\_food'', it transitions its state to three cases: 
% \begin{itemize}
%     \item If the user utters a food name (\{food\}), the system transitions its state to ``dig\_food'' and memorizes \{food\}.
%     \item If the user does not utter a food name and does not show ``negative'', the system transition its state to ``commenting\_food.''
%     \item If the user does not utter a food name and shows ``negative'', the system transition its state to ``just\_backchannel.''
% \end{itemize}

\section{Implementations}
\subsection{Four dialogue phases and recommendation}
Our dialogue flow consists of the following four phases.
The phases were designed on the basis of the system by Team kasuga, which participated in the past Dialogue Robot Competition \cite{IEEEref:robotcompe1} and was developed by a member of our laboratory.
% The phases were designed on the basis of the system by Team kasuga, which participated in the previous year's dialogue robot competition \cite{IEEEref:robotcompe_1} and was developed by a member of our laboratory.
OSbot asks about the following contents in each phase and talks with the user:
\begin{itemize}
    \item Opening: the user's name, hometown, travel destination, and transportation.
    \item Introduction of sightseeing spots: 
    impressions of sightseeing spots. OSbot then introduces the two spots the user selected.
    \item Questions about travel preferences: what the user usually eats when going out and the user's preference toward sightseeing destinations.
    \item Recommendation: OSbot recommends one sightseeing spot and gives some reasons for the recommendation.
\end{itemize}

In the phase in which OSbot asks the user questions (Opening, Introduction of sightseeing spots, Questions about travel preferences), OSbot works to prevent unexpected user utterances.
OSbot has two strategies to prevent a dialogue breakdown: it confirms the user's name after recognized it and utters sentences that are less dependent on the user's utterances \cite{IEEEref:dialogue_design}.
% It confirms the user's name after hearing the name, and it utters sentences that are less dependent on the user's utterances. 

In the Recommendation phase, OSbot uses the keywords obtained in the ``Question about travel preferences'' phase to show the reasons for the recommendation. 
It also introduces three places near the sightseeing spot that it recommends.
The place names were prepared beforehand.
% The places were prepared beforehand from Wikipedia\footnote{\url{https://ja.wikipedia.org}} and DBpedia\footnote{\url{https://ja.dbpedia.org/}}.
% We collected data by getting spot information that can be accessed from the page of a prefecture's sightseeing spots (for example, the page named ``Tokyo's sightseeing spots'').
% Unsuitable data were manually excluded.

\subsection{Guiding user utterances}
We make OSbot show an example answer when asking questions as a form of self-disclosure. 
OSbot uses self-disclosure to alleviate a user's confusion about what to answer.
For example, ``Hokkaido is memorable for me.
Do you have any recommendations for sightseeing spots?''

\subsection{Repeating keywords}
We use keywords extracted from a user utterance in the OSbot utterance that follows \cite{IEEEref:dialogue_design}.  
That is, OSbot's utterances include what the user said, e.g., ``noodles'' in Fig.~\ref{fig:dialogue_flow}.
This would have the effect of making the user feel that the system accepts and understands what he/she has said.

\subsection{Controlling voice and body posture}
\begin{figure*}[t]
  \centering
  \includegraphics[width=15cm]{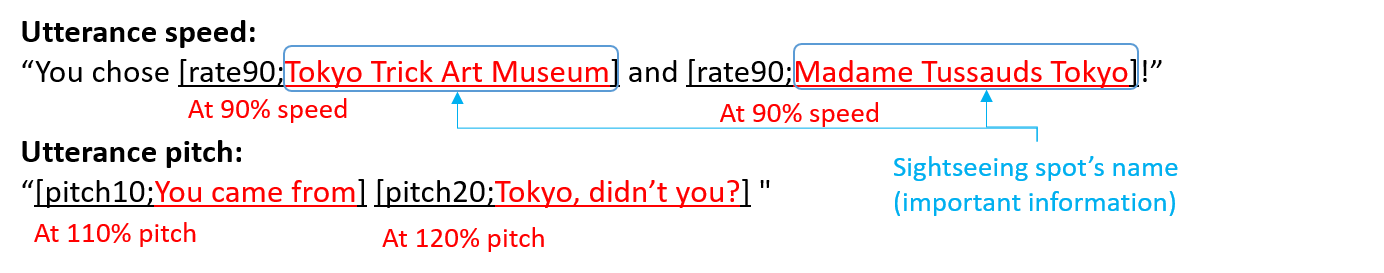}
  \caption{
  Examples of changing speed and pitch of speech.
  }
  \label{fig:sys_voice}
\end{figure*}
OSbot controls the speed and pitch of the robot's synthesized speech.
Example descriptions are shown in Fig.~\ref{fig:sys_voice}.
OSbot lowers the utterance speed to prevent a user from failing to catch important information, such as a sightseeing spot's name (which is often an unfamiliar proper noun) and the robot's name (Shoko).
The upper row in Fig.~\ref{fig:sys_voice} shows an example of lowering the robot's speaking rate to 90\% for the text surrounded in brackets
and then emphasizing the sightseeing spot names ``Tokyo Trick Art Museum'' and ``Madame Tussauds Tokyo.''
Furthermore, OSbot raises its voice pitch in steps to express joy naturally.
The lower row shows an example of raising the voice pitch to 110\% and 120\%, respectively, for each piece of text surrounded in brackets, and then OSbot expresses its pleasure from listening to the user's answer.

OSbot controls the robot's facial expression and body posture by sending commands in order to its pleasure.
We aim for a natural facial expression by combining multiple degrees of leaning forward and multiple types of smiling.
OSbot expresses its pleasure in order to make the user feel pleasure.

\subsection{Testing functions}
We used a logging function to test user utterance sequences when testing OSbot.
The log obtained from the logging function contains four types of information: a system utterance, a user utterance, a sentiment analysis result, and a current state name.
We also prepared test sequences of user utterances.
By using these functions and sequences, we checked dialogues efficiently.

We checked the generated dialogue logs several times.
In particular, we carefully checked the Japanese honorific expressions from the perspective of whether they were appropriate for a travel agent.
DialBB enables us to quickly check and edit the system utterances, the state transitions, and the transition conditions because it manages them in the form of a spreadsheet.

\section{Results}
\subsection{Dialogue examples}

We show two excerpts of dialogues conducted by our system.  
Labels starting with S or U in the leftmost column denote system or user utterances, respectively.

\begin{figure}[t]
\centering
\footnotesize
\begin{tabular}{lll}
   & Utterance                               & Sentiment \\\hline \hline
S1 & What do you usually eat when going out? & -         \\
U1 & Nothing special.                        & Negative  \\
S2 & Okay.                                   & -        
\end{tabular}
\caption{
Example dialogue when sentiment analysis was successful.
}
\label{fig:example_1}
\end{figure}
Fig.~\ref{fig:example_1} shows an example where the sentiment analysis result worked successfully.
The user replied with ``Nothing special,'' after OSbot asked the user about daily food.
The sentiment analyzer classified the user utterance as negative.
On the basis of the result, OSbot replied, ``Okay'' as S2, instead of commenting on the topic with ``I like eating \{food\} when going out too!'' and ``Eating delicious food makes us happy!''

\begin{figure}[t]
\centering
{\footnotesize
\begin{tabular}{lll}
   & Utterance                                                                                                                    & Keyword \\ \hline\hline
S1 & What do you usually eat when going out?                                                                                      & -       \\
U1 & I like noodles.                                                                                                              & noodles \\
S2 & \begin{tabular}[c]{@{}l@{}}I like eating noodles when going out too! \\Could you tell me about your favorite restaurant?\end{tabular}            & -       \\
U2 & \begin{tabular}[c]{@{}l@{}}Kourakuen's noodles are delicious. \\  (Kourakuen is a popular restaurant in Japan.)\end{tabular} & -       \\
S3 & Eating delicious food makes us happy!...                                                                                     & -      
\end{tabular}
}
\caption{
Example of dialogue when keyword extraction was successful.
}
\label{fig:example_2}
\end{figure}
Fig.~\ref{fig:example_2} shows an example of success in using keyword extraction.
In turn U1, the user answered ``I like noodles'' after OSbot asked about eating out, and the keyword  ``noodles'' was extracted from the user utterance.
Using the keyword extraction result, OSbot showed empathy like ``I like eating noodles when going out too!'' in the first part of S2, and then continued to dig by asking a question about the user's favorite restaurant in the latter part of S2.

\subsection{Scores in preliminary round}
\begin{figure*}[t]
{\scriptsize
\centering
\begin{tabular}{c|r|r|r|r|r|r|r|r|r|}
\cline{2-10}
                               & \multicolumn{1}{c|}{\begin{tabular}[c]{@{}c@{}}Satisfaction \\ with choice\end{tabular}} & \multicolumn{1}{c|}{Informativeness} & \multicolumn{1}{c|}{Naturalness} & \multicolumn{1}{c|}{Appropriateness} & \multicolumn{1}{c|}{Likeability} & \multicolumn{1}{c|}{\begin{tabular}[c]{@{}c@{}}Satisfaction \\ with dialogue\end{tabular}} & \multicolumn{1}{c|}{Trustworthiness} & \multicolumn{1}{c|}{Usefulness} & \multicolumn{1}{c|}{\begin{tabular}[c]{@{}c@{}}Intention \\ to reuse\end{tabular}} \\ \hline
\multicolumn{1}{|c|}{Average}  & 5.16                                                                                     & 4.96                                 & 4.36                             & 5.16                                 & 5.04                             & 5.16                                                                                       & 5.04                                 & 4.84                            & 4.92                                                                               \\ \hline
\end{tabular}
}
\caption{Averaged scores of each question. Description for each question is written in \cite{IEEEref:robotcompe_2022}.}
\label{fig:whole_score}
\end{figure*}

Two kinds of questionnaire items were conducted in the preliminary round: preference between two sightseeing spots and dialogue quality.
The former involved the degree of the preference after the system recommended the spot.
The increase in the degree was 18.12, and this score was the third highest.
The latter questionnaire items were about the dialogue quality, each of which was measured on a 7-point Likert scale \cite{IEEEref:robotcompe_2022}.
The averaged scores for each question are shown in Fig.~\ref{fig:whole_score} and their average was 4.96.

The averaged scores were high in appropriateness and satisfaction with the dialogue because OSbot shows appropriate reactions to users' utterances, like those in Figs.~\ref{fig:example_1}~and~\ref{fig:example_2}.
Showing more convincing information on sightseeing spots would improve the scores in informativeness and robot recommendation effect.

\section{Conclusion}
We explained our dialogue robot system, OSbot, developed for Dialogue Robot Competition 2022.
OSbot determines next system utterances on the basis of sentiment analysis and keyword extraction.
It also controls the speed of synthesized speech so that users do not miss important words as well as voice pitch, body posture, and facial expression to show that the robot feels happy to hear the user's utterance.
Use of DialBB enabled us to describe and check dialogue flow and system utterances efficiently.
As a result, OSbot passed the preliminary round of Dialogue Robot Competition 2022 and placed third.

\section*{ACKNOWLEDGEMENT}
This work was partly supported by JSPS KAKENHI Grant Number JP19H05692.

\bibliographystyle{IEEEtran}
\bibliography{IEEEabrv, IEEEref}

\begin{thebibliography}{1}
\providecommand{\url}[1]{#1}
\csname url@rmstyle\endcsname
\providecommand{\newblock}{\relax}
\providecommand{\bibinfo}[2]{#2}
\providecommand\BIBentrySTDinterwordspacing{\spaceskip=0pt\relax}
\providecommand\BIBentryALTinterwordstretchfactor{4}
\providecommand\BIBentryALTinterwordspacing{\spaceskip=\fontdimen2\font plus
\BIBentryALTinterwordstretchfactor\fontdimen3\font minus
  \fontdimen4\font\relax}
\providecommand\BIBforeignlanguage[2]{{%
\expandafter\ifx\csname l@#1\endcsname\relax
\typeout{** WARNING: IEEEtran.bst: No hyphenation pattern has been}%
\typeout{** loaded for the language `#1'. Using the pattern for}%
\typeout{** the default language instead.}%
\else
\language=\csname l@#1\endcsname
\fi
#2}}

\bibitem{IEEEref:robotcompe_2022}
T.~Minato, R.~Higashinaka, H.~Nishizaki, T.~Nagai, K.~Sakai, and T.~Funayama,
  ``Overview of dialogue robot competition 2022,'' in \emph{Proc. of the
  Dialogue Robot Competition 2022}, 2022.

\bibitem{IEEEref:robotcompe_robot}
R.~Higashinaka, T.~Minato, K.~Sakai, T.~Funayama, H.~Nishizaki, and T.~Nagai,
  ``Spoken dialogue system development at the dialogue robot competition,''
  \emph{The Journal of the Acoustical Society of Japan}, vol.~77, no.~8, pp.
  512--520, 2021.

\bibitem{IEEEref:robotcompe1}
R.~Higashinaka, T.~Minato, K.~Sakai, T.~Funayama, H.~Nishizaki, and T.~Nagai,
  ``Dialogue robot competition for the development of an android robot with
  hospitality,'' in \emph{Proc. of GCCE2022}, 2022.

\bibitem{IEEEref:w2v}
T.~Mikolov, K.~Chen, G.~Corrado, and J.~Dean, ``Efficient estimation of word
  representations in vector space,'' in \emph{Proc. of Workshop at ICLR}, 2013.

\bibitem{komatani_hazumi}
K.~Komatani and S.~Okada, ``Multimodal human-agent dialogue corpus with
  annotations at utterance and dialogue levels,'' in \emph{Proc. of
  International Conference on Affective Computing and Intelligent Interaction
  (ACII)}, 2021, pp. 1--8.

\bibitem{IEEEref:bert}
\BIBentryALTinterwordspacing
N.~Reimers and I.~Gurevych, ``Sentence-bert: Sentence embeddings using siamese
  bert-networks,'' in \emph{Proc. of the 2019 Conference on Empirical Methods
  in Natural Language Processing}.\hskip 1em plus 0.5em minus 0.4em\relax
  Association for Computational Linguistics, 11 2019. [Online]. Available:
  \url{http://arxiv.org/abs/1908.10084}
\BIBentrySTDinterwordspacing

\bibitem{IEEEref:coucke2018snips}
A.~Coucke, A.~Saade, A.~Ball, T.~Bluche, A.~Caulier, D.~Leroy, C.~Doumouro,
  T.~Gisselbrecht, F.~Caltagirone, T.~Lavril, M.~Primet, and J.~Dureau, ``Snips
  voice platform: an embedded spoken language understanding system for
  private-by-design voice interfaces,'' \emph{arXiv preprint arXiv:1805.10190},
  pp. 12--16, 2018.

\bibitem{IEEEref:dialogue_design}
K.~Komatani, R.~Takeda, K.~Nakashima, and M.~Nakano, ``Design guidelines for
  developing systems for dialogue system competitions,'' in \emph{Proc. of
  International Workshop on Spoken Dialogue System Technology (IWSDS)},
  vol.~16, 2021.

\end{thebibliography}

\end{document}